\def\tdot{\stackrel{\dots}}
\def\order{{\cal O}}
\documentstyle[pre,twocolumn,aps,psfig]{revtex}
\begin{document}
\preprint{IASSNS-HEP-99/41}
\draft
\twocolumn[\hsize\textwidth\columnwidth\hsize\csname@twocolumnfalse%
\endcsname
\title{The Radiative Kicked Oscillator: A Stochastic Web or Chaotic
Attractor ?}
\author{
Y. Ashkenazy$^{\text{1}}$\cite{addressA} and
L. P. Horwitz$^{\text{2}}$\cite{addressB}}
\address{$^1$ Department of Physics, Bar-Ilan University, Ramat-Gan 52900,
Israel\\
$^2$ Institute for Advanced Study, Princeton, N.J. 08540.}
\date{\today}
\maketitle
\begin{abstract}
{
A relativistic charged particle moving in a uniform magnetic field and kicked 
by an
electric field is considered. Under the assumption of small magnetic field, an
iterative map is developed.  We consider both the case in which no
radiation is assumed and the radiative case, using the Lorentz-Dirac
equation to describe the motion.
Comparison between the  non-radiative
case and the radiative case shows that in both cases one can
observe
a stochastic web structure for weak magnetic fields, and, although
there are global differences in the result of the map, that both cases
are qualitatively similar in their small scale behavior.
We also develop an iterative map for strong magnetic fields.
In that case the web structure  no longer exists; it is replaced by a rich
chaotic behavior. It is shown that the particle does not diffuse to infinite
energy; it is limited by the boundaries of an attractor (the boundaries are
generally much smaller than light velocity). Bifurcation occurs,
converging rapidly to Feigenbaum's universal constant.  The chaotic
behavior appears to be robust.
 For intermediate magnetic fields,
it is more difficult to observe the web structure, and the influence of the
unstable fixed point is weaker.
}
\end{abstract}
]
\narrowtext
\newpage

\section{Introduction}

Zaslavskii {\it et al} \cite{Zaslavskii86} studied the behavior of particles
in the wave packet of an electric field in the presence of a
static magnetic field.
For a broad wave packet with sufficiently uniform spectrum, one may show that
the problem can be stated in terms of an electrically kicked harmonic
oscillator. For rational ratios between the frequency
of the kicking field and the Larmor frequency associated with the
magnetic field the phase space of the system is covered by a mesh of finite
thickness; inside the filaments of the mesh, the dynamics of the particle is
stochastic and
outside (in the cells of stability), the dynamics is regular. This structure is
called a stochastic web. It was found that this pattern covers the entire
phase plane, permitting the particle to diffuse arbitrarily far into the
region of high energies (a process analogous to Arnol'd
diffusion \cite{Arnold64}).

Since the stochastic web leads to unbounded energies, several authors have
considered the corresponding relativistic problem (for a work which can be
related to the relativistic stochastic web see \cite{Chernikov89}).
Longcope and Sudan
\cite{Longcope87} studied this system (in effectively
$1\frac{1}{2}$ dimensions) and found that for initial conditions close to the
origin of the phase space there is a stochastic web, which is bounded in
energy, of a form quite similar, in the neighborhood of the origin, to the
non-relativistic case treated by
Zaslavskii {\it et al}. Karimabadi and Angelopoulos \cite{Karimabadi89}
studied the case of an obliquely propagating wave, and showed that under
certain conditions, particles can be accelerated to unlimited energy through
an Arnol'd diffusion in two dimensions.
Since an accelerated charged particle radiates, it is
important to study the radiative corrections to this motion. We shall use the
Lorentz-Dirac equation to compute this effect.

We compute solutions to this equation for the
case of the kicked oscillator. Under the restriction of weak magnetic field,
at low velocities, the stochastic web
found by Zaslavskii {\it et al}\cite{Zaslavskii86} occurs; the system diffuses
in the stochastic region to unbounded energy, as found by Karimabadi and
Angelopoulos\cite{Karimabadi89}. The velocity of the particle is light
speed limited by the dynamical equations, in particular, by the suppression
of the action of the electric field at velocities approaching the velocity
of light \cite{Goldman}.

For the case of a strong magnetic field case 
(a case which can not be examined in
laboratory) the stochastic web, which may occur in the weak magnetic
field case, does not appear and is replaced by a rich chaotic
behavior. In this regime, the particle does not accelerate to
infinite energy; it limited by the boundary of the chaotic attractor.

\section{Model}

In the present study we will consider a charged particle moving in a uniform
magnetic field, and kicked by an electric field. The effect of relativity, as
well
as the radiation of the particle, will be considered. We restrict ourselves to
the ``on mass shell'' constraint (keeping constant ${\dot x}^\mu{\dot
x}_\mu = -c^2$)\footnote{The more general case which includes also
the possibility ``off mass shell'' motion, is discussed in Ref.
\cite{Horwitz}.}.

The fundamental equation that we use to study radiation is the Lorentz-Dirac
equation \cite{Dirac},
\begin{equation}
\label{e1}
m_0\ddot x^\mu={e \over c}\dot x_\nu F^{\mu\nu} + \gamma_0m_0({\tdot x^\mu}-
{1 \over c^2} \dot x^\mu \ddot x_\nu \ddot x^\nu),
\end{equation}
where $\gamma_0={2 \over 3}{r_0 \over c}=6.26\times 10^{-24}\sec$. The
indices $\mu$ and
$\nu$ indicate the coordinates $t$, $x$, $y$, and $z$ (or 0, 1, 2, 3),
and the derivative is
with respect to $\tau$, which can be regarded as proper time in the
``on-mass-shell'' case. $F^{\mu\nu}$ is the antisymmetric electromagnetic
tensor. The first term of the right hand side of Eq. (\ref{e1}) is the
relativistic Lorentz force, and the second term is the radiation-reaction
term. Note that the small size of the radiation coefficient
($\gamma_0 \ll 1$) leads to a singular equation that requires special
mathematical treatment, as well as physical restrictions, 
as will be discussed in the succeeding sections.

Following Zaslavskii {\it et al} \cite{Zaslavskii86}, the magnetic field is
chosen to be  uniform  in the $z$ direction, and the kicking
electric field is a function of $x$ in the $x$ direction,
\begin {eqnarray}
\label{e2}
{\bf B} &=& (0,0,B) \nonumber \\
{\bf E}(x,t) &=& ({f(x){\sum_{n=-\infty}^\infty\delta(t-nT)}},0,0).
\end{eqnarray}
Originally, Zaslavskii {\it et al} chose a uniform
broad band electric field wave packet which can be expanded as an infinite sum
of (kicking) $\delta$-functions ($\omega_0$ is the frequency of the central
harmonic of the wave packet, $k_0$ is the wave number of the central harmonic,
and $\Delta \omega$ is the frequency between the harmonics of the wave packet)
\begin{equation}
\label{e3a}
E_x=E(x,t)=-E_0\sum_{n=-\infty}^\infty \sin (k_0x-\omega_0t-n\Delta\omega t),
\end{equation}
which, for $\omega_0=0$, becomes
\begin{equation}
\label{e3}
E(x,t)= -E_0T\sin(k_0x)\sum_{n=-\infty}^\infty\delta(t-nT),
\end{equation}
where $T={{2\pi}\over{\Delta\omega}}$.
Eq. (\ref{e2}) is the generalization of Eq. (\ref{e3}), with an
arbitrary function $f(x)$ instead of the sine function of Eq. (\ref{e3}).

In order to introduce a map which connects between cycles of integration
which start before a kick and end before the next electric field
kick, one has first to integrate over the $\delta$ electric kick and then
to integrate the equations of motion between the kicks, where only a uniform
magnetic field is present and there is no electric field up to the beginning
of the next kick.

\section{The motion of a charged particle in a uniform magnetic field}

\subsection{An integration with respect to the proper time $\tau$}

In the case of a uniform magnetic field in Eq. (\ref{e2}), Eqs. (\ref{e1})
reduce to three coupled differential equations,
\begin{eqnarray}
\label{e4}
c \ddot t &=\ddot x_0=& \gamma_0({\tdot x_0} - {1 \over {c^2}}
\dot x_0R) \nonumber \\
\ddot x &=\ddot x_1=& -\Omega \dot x_2 + \gamma_0({\tdot x_1} -
{1 \over {c^2}} \dot x_1R) \\
\ddot y &=\ddot x_2=& \Omega \dot x_1 + \gamma_0({\tdot x_2} -
{1 \over {c^2}} \dot x_2R), \nonumber
\end{eqnarray}
where $\Omega={{e_0B_0} \over {m_0c}}$,
$R=\ddot x_1^2+\ddot x_2^2-\ddot x_0^2$, and $e_0=-e$ (charge of the electron).

Using a complex coordinate\cite{Sokolov}, $u=x+iy$, Eqs. (\ref{e4}) can be
written as,
\begin{eqnarray}
\label{e5}
\ddot t &=& \gamma_0({\tdot t}-{1 \over {c^2}} \dot t R) \nonumber \\
\ddot u &=& i\Omega\dot u+\gamma_0({\tdot u}-{1 \over {c^2}} \dot u R).
\end{eqnarray}
 It is very convenient to use the hyperbolic coordinates,
\begin{eqnarray}
\label{e6}
\dot t &=& \cosh q \nonumber \\
\dot x &=& c \sinh q \cos \phi \\
\dot y &=& c \sinh q \sin \phi \nonumber
\end{eqnarray}
since the constraint (see \cite{Horwitz} for further
discussion),
\begin{equation}
\label{e7}
\dot x^\mu \dot x_\mu = \dot x^2+\dot y^2-c^2 \dot t^2 = -c^2,
\end{equation}
is then automatically satisfied. The complex coordinate $u$ then becomes,
\begin{equation}
\label{7a}
\dot u = c \sinh q e^{i\phi}
\end{equation}
and the equations of motion Eqs. (\ref{e5}) can be written as,
\begin{eqnarray}
\dot q &=& -\gamma_0 \dot\phi^2 \cosh q\sinh q + \gamma_0\ddot q \label{e8}\\
\dot \phi &=& \Omega +2\gamma_0\dot\phi\dot q \coth q+\gamma_0 \ddot \phi
\label{e9}.
\end{eqnarray}
As pointed out earlier, in the case of a singular equation (such as Eqs.
(\ref{e8}) and (\ref{e9})), one has to consider physical arguments as well as 
mathematical arguments. There are several suggested methods to avoid the
``run away electron'' problem \cite{Dirac}. One of the most frequently used,
especially useful for scattering problems, assumes \cite{Rohrlich} that the
particle loses all its energy after a sufficiently large time, and the run
away parts of the solution are set to zero; the equations
 of motion can then be written as an integral equation. Another method,
which permits the study of problems with only non-asymptotic states
\cite{Aguirrebiria},
uses an iterative singular perturbation integration that leads to the stable
solution. Since the  integration we must use is over a finite time,
 it is
impossible to use an asymptotic condition (e.g. $|v| \to 0$ at $\tau \to
\infty$).
The approach of Sokolov and Ternov  \cite{Sokolov} is more suitable for our
purpose, since it can be implemented for bounded time integration, and the
mathematical solution is simple. In this method, in the first step,
the perturbation terms in Eq. (\ref{e9}) are disregarded (as in the iterative
scheme of \cite{Aguirrebiria}), and then the resulting
equation is exactly integrated. In the second step, the singular term on
the right hand side of
Eq. (\ref{e8}) is not considered; using $\phi (\tau)$ from Eq. (\ref{e9}), the
equation including just the first term is integrated. The solution is
\cite{Sokolov}
\begin{eqnarray}
\label{e10}
\phi &=& \Omega \tau \nonumber \\
\beta(\tau) &=& \tanh q = {v_0 \over c} e^{-\gamma_0\Omega^2 \tau} =
\beta_0 e^{-{\tau \over \tau_0}},
\end{eqnarray}
where $\beta_0={v_0 \over c}$ is the actual initial velocity divided
by $c$, and
$\tau_0={1 \over {\gamma_0\Omega^2}}$ is the decay time for the energy of the 
particle.

To get an estimate for the radiation during one cycle, consider
 the maximal uniform magnetic field that can achieved today in a
laboratory, which is\footnote{It
is possible to obtain a short-lived magnetic field of
$\order(10^6\,{\rm gauss})$ in the laboratory.}
$\order(10T) = \order(10^5\, {\rm gauss})$. If, for example,
$B_0=10^5\, {\rm gauss}$, then $\Omega=1.76\times10^{12}{1 \over \sec}$, and
$\tau_0=5\times 10^{-2} \sec$. Thus, it is clear from Eq. (\ref{e10}) that the
particle makes ${\Omega \over {2\pi}}\tau_0 \approx 10^{10}$ cycles before
it decays to ${1 \over e}$ of its initial velocity. In other words, the energy
loss during one cycle in very small, and since in our problem, the time $T$
between the kicking is of the order of the period, the energy loss
between consecutive kicks is very small.

\begin{figure}
\psfig{figure=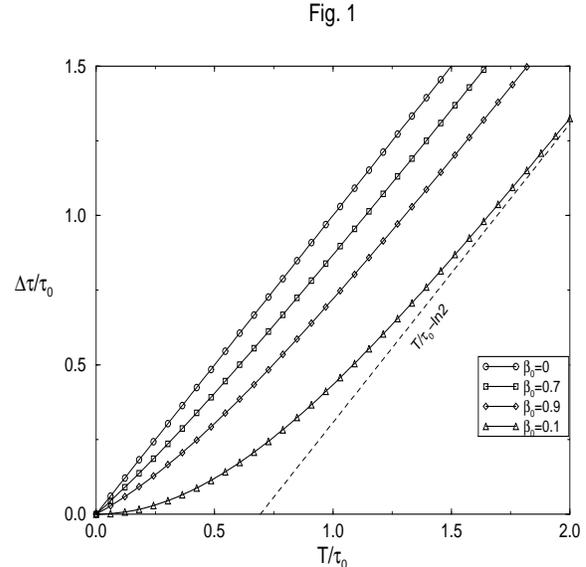,height=8cm,width=8.5cm,angle=-90}
\caption[]{\label{fig1}
The function ${{\Delta \tau (T / \tau_0)}\over{\tau_0}}$ versus
$T / \tau_0$. Four typical
cases are shown ($\beta_0=0$, $\beta_0=0.7$, $\beta_0=0.9$, $\beta_0=1$).
All possible curve lines between the graphs of $\beta_0=0$ and $\beta_0=1$.
The dashed line indicates the asymptotic behavior of the case $\beta_0=1$.
}
\end{figure}

The time between the kicking is measured according to the observed time along 
the motion,
$\Delta t = T$. Thus, it is necessary to find the corresponding $\Delta \tau$.
It follows from Eq. (\ref{e6}) and Eq. (\ref{e10}) that,
\begin{equation}
\label{e11}
\dot t = \cosh q = {1 \over \sqrt{1-\beta^2_0 e^{-{{2\tau} \over {\tau_0}}}}}.
\end{equation}
The solution of Eq. (\ref{e11}) can be obtained by a elementary integration,
\begin{equation}
\label{e12}
t=\tau_0\ln \left({{1+\sqrt{1-a^2}} \over a}C_1\right),
\end{equation}
where $a =\beta_0 e^{-{\tau \over {\tau_0}}}$. After some algebra
 one gets,
\begin{equation}
\label{e13}
\Delta \tau = \tau_0 \ln \left( {1 \over 2}{{\beta_0^2+(1+\sqrt{1-\beta_0^2})^2
e^{2T/\tau_0}} \over {(1+\sqrt{1-\beta_0^2})e^{T/\tau_0}}} \right).
\end{equation}
In Fig. \ref{fig1} we present the behavior of the function
${{\Delta \tau (T / \tau_0)}\over{\tau_0}}$. It can be seen from Fig.
\ref{fig1} that in any case,
\begin{equation}
\label{e14}
\Delta \tau \le T;
\end{equation}
this implies  that the time difference according to the
 proper time $\tau$ is always less then the time difference according
to observed time $t$. This fact is corresponds to the well known relativistic
``time-dilation'' phenomenon. The inequality Eq. (\ref{e14})
can be also illustrated by considering the two extreme cases,
\begin{eqnarray}
T \ll \tau_0 & \Rightarrow & \Delta \tau \approx T\sqrt{1-\beta_0^2}
+{{\beta_0^2} \over 2}{{T^2} \over {\tau_0}}+ \order(T^3) \label{e16}\\
T \gg \tau_0 & \Rightarrow & \Delta \tau \approx T+\tau_0 \ln
\left({{1+\sqrt{1-\beta_0^2}} \over 2} \right) \label{e15}.
\end{eqnarray}
For strong magnetic field (i.e. the case of Eq. (\ref{e15})) the numerator is
less than (or equal) to 2 and thus the $ln$ function will give a negative 
number; in this case the inequality Eq. (\ref{e14}) is clearly achieved.
For a low magnetic field Eq. (\ref{e16}) behaves like a parabola;
the slope at $T=0$
is $\sqrt{1-\beta_0^2}$ which is less than one, and thus the inequality Eq.
(\ref{e14}) is satisfied.
In the present study we will consider separately the cases of weak magnetic
field (Eq. (\ref{e16}) and strong magnetic field (Eq. (\ref{e15})).

\subsection{An integration with respect to the observed time $t$}

In the previous subsection we have shown the approximate solution for a charged
particle moving in a uniform magnetic field. However, as mentioned above,
 the kicking of
the electric field from Eq. (\ref{e2}) is with respect to the observed time
$t$, and it is therefore convenient to integrate the equations of motion
(Eqs. (\ref{e8}) and (\ref{e9})) with respect to $t$.

The arguments that were used in order to obtain Eqs. (\ref{e10}),
as well as the implementation of the chain rule on Eqs. (\ref{e8}) and
(\ref{e9}) by the use of Eq. (\ref{e6}), lead to the following equations of
motion
\begin{eqnarray}
{{dq}\over{dt}} &=& -{1\over{\tau_0}}\sinh q \label{e17} \\
{{d\phi}\over{dt}} &=& {\Omega \over {\cosh q}} \label{e18}.
\end{eqnarray}
A simple integration of Eq. (\ref{e17}) yields,
\begin{equation}
\label{e19}
\tanh {q\over 2} = e^{C_1}e^{-{t \over{\tau_0}}},
\end{equation}
where $C_1<0$.
Substitution of Eq. (\ref{e19}) in Eq. (\ref{e18}) gives,
\begin{equation}
\label{e20}
{{d\phi}\over{dt}} = \Omega \tanh ({t\over{\tau_0}}-C_1),
\end{equation}
and the solution is
\begin{equation}
\label{e21}
\phi = \Omega\tau_0\ln\cosh({t\over {\tau}}-C_1)+C_2,
\end{equation}
where $C_1=\cosh^{-1}({1\over{\beta_0}})$, and $C_2=\tan^{-1}
(\dot y_0 / \dot x_0)+\Omega\tau_0\ln\beta_0$.
Returning to the original $x$, $y$ coordinates (using Eqs. (\ref{e6})),
the equations of motion become,
\begin{eqnarray}
{{dx} \over {dt}} &=& c {{\cos(\Omega\tau_0\ln(\cosh(t/\tau_0-C_1))+C_2)} \over
{\cosh(t/\tau_0-C_1)}} \nonumber \\
{{dy} \over {dt}} &=& c {{\sin(\Omega\tau_0\ln(\cosh(t/\tau_0-C_1))+C_2)} \over
{\cosh(t/\tau_0-C_1)}} \label{e22}.
\end{eqnarray}
It is possible to write Eq. (\ref{e22}) in more convenient way, by using
the relation
\begin{equation}
\label{e23}
\cosh({t \over {\tau_0}}-C_1)={1 \over {\beta_0}} \cosh{t \over {\tau_0}}
+{1 \over {\beta_0}}\sqrt{1-\beta_0^2}\sinh{t \over {\tau_0}}.
\end{equation}
Eqs. (\ref{e22}) then become
\begin{eqnarray}
{{dx} \over {dt}} &=& {{\left({{dx}\over{dt}}\right)_0\cos\alpha-
\left({{dy}\over{dt}}\right)_0\sin\alpha} \over {\cosh{t\over{\tau_0}}+
\sqrt{1-\beta_0^2}\sinh{t\over{\tau_0}}}} \nonumber \\
{{dy} \over {dt}} &=& {{\left({{dx}\over{dt}}\right)_0\sin\alpha+
\left({{dy}\over{dt}}\right)_0\cos\alpha} \over {\cosh{t\over{\tau_0}}+
\sqrt{1-\beta_0^2}\sinh{t\over{\tau_0}}}} \label{e24},
\end{eqnarray}
where
\begin{equation}
\label{e25}
\alpha = \Omega\tau_0\ln (\cosh{t\over{\tau_0}}(1+\sqrt{1-\beta_0^2}
\tanh{t\over{\tau_0}})).
\end{equation}

It is necessary to integrate Eq. (\ref{e24}) since the value of $x$ is used in
the electric field kicking in Eq. (\ref{e2}). There is no analytical
solution to Eq. (\ref{e24}), and a numerical integration must, in general,
be performed. In the succeeding sections we will study separately the case of
weak magnetic field ({\it i.e.} $T/\tau_0\ll 1$) and the case of strong
magnetic field ({\it i.e.} $T/\tau_0\gg 1$).

\section{Derivation of the map - weak magnetic field}

\subsection{The approximated solution for the weak magnetic field case}

When dealing with weak magnetic field we restrict ourselves to
$T/\tau_0\ll 1$, and it is possible to
expand ${{dx}\over{dt}}$ and ${{dy}\over{dt}}$ in a Taylor series and then
to integrate. The expansion of Eq. (\ref{e25}) is
\begin{equation}
\label{e26}
\alpha \approx \Omega \sqrt{1-\beta_0^2}t
+{1 \over 2}\beta_0^2{{t^2} \over {\tau_0}}+{1\over 3}\sqrt{1-\beta_0^2}
{{t^3} \over {\tau_0^2}}({3\over 2}-\beta_0^2)+
\order({{{t^4} \over {\tau_0^3}}}).
\end{equation}
The first order expansion (according to $t/\tau_0$) of Eqs. (\ref{e24}) is
\begin{eqnarray}
{{dx} \over {dt}} = {1\over{1+{1\over\gamma}{t\over{\tau_0}}}} &&
\left[\left({{dx}\over{dt}}\right)_0\cos({{\Omega}\over\gamma}t+{1\over2}
\Omega\beta_0^2{{t^2}\over{\tau_0}}) \right. \nonumber \\
&& \left. -\left({{dy}\over{dt}}\right)_0\sin({{\Omega}\over\gamma}t+{1\over2}
\Omega\beta_0^2{{t^2}\over{\tau_0}})\right] \nonumber \\
{{dy} \over {dt}} = {1\over{1+{1\over\gamma}{t\over{\tau_0}}}} &&
\left[\left({{dx}\over{dt}}\right)_0\sin({{\Omega}\over\gamma}t+{1\over2}
\Omega\beta_0^2{{t^2}\over{\tau_0}})  \right. \nonumber \\
&& \left. +\left({{dy}\over{dt}}\right)_0\cos({{\Omega}\over\gamma}t+{1\over2}
\Omega\beta_0^2{{t^2}\over{\tau_0}})\right], \label{e27}
\end{eqnarray}
where $\gamma=1/\sqrt{1-\beta_0^2}$.
Since ${1\over\gamma}{t\over{\tau_0}}\ll 1$, one can use the approximation,
\begin{equation}
\label{e28}
{1\over{1+{1\over\gamma}{t\over{\tau_0}}}} \approx
{{1-{1\over\gamma}{t\over{\tau_0}}}}.
\end{equation}
Using Eq. (\ref{e28}), the actual velocities, $dx/dt$ and $dy/dt$, from Eq.
(\ref{e27}) can be integrated and expressed by elementary Fresnel functions.
However, the effect of radiation is due to the ${1\over\gamma}{t\over{\tau_0}}$
term (which multiplies the sine and cosine functions), and the
${1\over2}\Omega\beta_0^2{{t^2}\over{\tau_0}}$ term is not essential since the
major offset from the unstable fixed point is due to the
${{\Omega}\over\gamma}t$ term.

Under the above assumptions the resulting equations are,
\begin{eqnarray} 
\left({{dx} \over {dt}}\right)_T &=& \left(1-{T\over{\gamma\tau_0}}\right) 
\times \nonumber \\
&& \left[\left({{dx}\over{dt}}\right)_0\cos({{\Omega}\over\gamma}T)-
\left({{dy}\over{dt}}\right)_0\sin({{\Omega}\over\gamma}T)\right] \label{e29}\\
\left({{dy} \over {dt}}\right)_T &=& \left(1-{T\over{\gamma\tau_0}}\right)
\times \nonumber \\
&& \left[\left({{dx}\over{dt}}\right)_0\sin({{\Omega}\over\gamma}T)+
\left({{dy}\over{dt}}\right)_0\cos({{\Omega}\over\gamma}T)\right]. \label{e30}
\end{eqnarray}
One therefore obtains
\begin{eqnarray}
x_T \approx {\gamma \over \Omega}&&\left[\left({{dy}\over{dt}}\right)_T
-\left({{dy}\over{dt}}\right)_0\right]- \nonumber \\
&& {\gamma \over {\Omega^2\tau_0}}\left[\left({{dx}\over{dt}}\right)_T
-\left({{dx}\over{dt}}\right)_0\right]+x_0. \label{e31}
\end{eqnarray}
The exponential decay from Eqs. (\ref{e22}) is replaced by linear decay; by 
squaring Eq. (\ref{e29}) and Eq. (\ref{e30}) one obtains the linear relation
$v(T)=\sqrt{v_x^2+v_y^2}=v_0(1-T/\gamma\tau_0)$.

\subsection{Integration over a $\delta$ electric kick}

As pointed out in the previous section, the Lorentz-Dirac equation
(Eq. (\ref{e1}))
is a singular equation. The electric field which is used in this paper was
expanded to a sum of $\delta$ functions (Eq. (\ref{e2})). In that case, Eq.
(\ref{e1}) can not be integrated over the kick by regular treatment, and some
approximation
for the $\delta$ function should be considered instead. However, since the
radiation has its effect in the direction of the relevant coordinate 
(in our case
it is the $x^1=x$ coordinate), and because of the  infinitesimal time
interval, it is possible to replace the kicking parameter, $K$, by an effective
kicking strength, which should be a smaller number than the original one.
Thus, the {\it nature} of the solution for the
system will not change due to the radiation during the kicking. The constant
magnetic field during the kick is not considered since the integration is over
an infinitesimal time interval.

Under the above assumptions, in the neighborhood of the kick,
the Lorentz-Dirac equation, Eq. (\ref{e1}), can
be written as,
\begin{eqnarray}
{d\over{dt}}\left({{dx}\over{d\tau}}\right) &=&
{f(x){\sum_{n=-\infty}^\infty\delta(t-nT)}} \nonumber \\
{d\over{dt}}\left({{dy}\over{d\tau}}\right) &=& 0.\label{e32}
\end{eqnarray}
Integration over the $\delta$ function yields,
\begin{eqnarray}
\left({{dx}\over{d\tau}}\right)_+ &=& \left({{dx}\over{d\tau}}\right)_-
+ f(x) \nonumber \\
\left({{dy}\over{d\tau}}\right)_+ &=& \left({{dy}\over{d\tau}}\right)_-
\label{e33}
\end{eqnarray}
where the + sign indicates after the kick, and the - sign, before the kick.
Following Ref. \cite{Zaslavskii86}, we choose,
\begin{equation}
\label{e34}
f(x) = {{eET}\over{m_0}}\sin\,kx = K\sin\,kx.
\end{equation}
Using the fact that ${d\over{d\tau}}=\gamma{d\over{dt}}$ one obtains the
following relations,
\begin{eqnarray}
\gamma_+ &=& \sqrt{1+{1\over{c^2}}\left[\left(\gamma_-\left({{dx}\over{dt}}
\right)_-+f(x_n)\right)^2+\gamma_-^2\left({{dy}\over{dt}}\right)_-^2\right]}
\nonumber \\
&&\left({{dx}\over{dt}}\right)_+ = {\gamma_-\over\gamma_+}\left({{dx}\over{dt}}
\right)_-+{1\over\gamma_+}f(x_n) \label{e35} \\
&&\left({{dy}\over{dt}}\right)_+ = {\gamma_-\over\gamma_+}\left({{dy}\over{dt}}
\right)_-\nonumber.
\end{eqnarray}
Returning to the initial charge $e=-e_0$ and thus $\Omega \to -\Omega$, the map
which connects the velocities from just before the kick to the next kick
(by the use of Eqs. (\ref{e29})-(\ref{e31})) is,
\begin{eqnarray} 
\left({{dx} \over {dt}}\right)_{n+1}&&=\left(1-{T\over{\gamma_+\tau_0}}\right)
\times \nonumber \\
&& \left[\left({{dx}\over{dt}}\right)_+\cos({{\Omega}\over{\gamma_+}}T)
+\left({{dy}\over{dt}}\right)_+\sin({{\Omega}\over{\gamma_+}}T)
\right] \label{e36}\\
\left({{dy} \over {dt}}\right)_{n+1}&&=\left(1-{T\over{\gamma_+\tau_0}}\right)
\times \nonumber \\
&& \left[-\left({{dx}\over{dt}}\right)_+\sin({{\Omega}\over{\gamma_+}}T)
+ \left({{dy}\over{dt}}\right)_+\cos({{\Omega}\over{\gamma_+}}T)
\right] \label{e37}\\
x_{n+1} &&\approx -{{\gamma_+} \over \Omega}\left[\left({{dy}\over{dt}}\right)_
{n+1}-\left({{dy}\over{dt}}\right)_+\right] \nonumber \\
&&-{{\gamma_+} \over {\Omega^2\tau_0}}\left[\left({{dx}\over{dt}}\right)_{n+1}
-\left({{dx}\over{dt}}\right)_+\right]+x_n. \label{e38}
\end{eqnarray}
Notice that the minus signs in Eq. (\ref{e35}) refer to the $n^{th}$ points of
the map.

It is possible to return to the non-radiative limit by letting
${1\over{\tau_0}}\to 0$. In that case the map is equivalent to the map which
was derived by Longcope and Sudan \cite{Longcope87}. The nonrelativistic limit,
which was derived by Zaslavskii {\it et al} \cite {Zaslavskii86},
is achieved by letting ${1\over{\tau_0}}\to 0$ and $c\to \infty$.

\subsection{Analysis} \label{s4c}

\begin{figure}
\psfig{figure=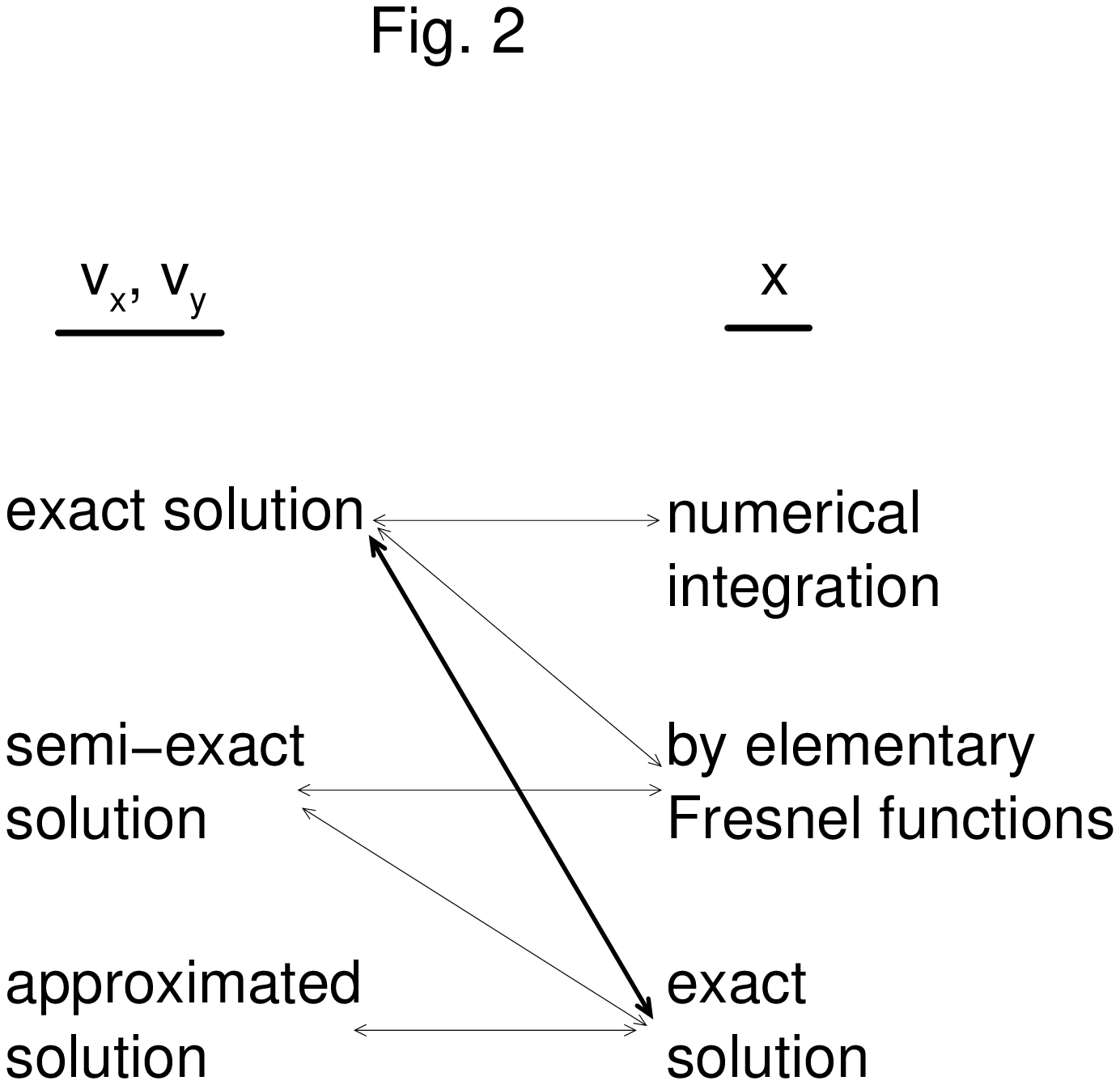,height=8cm,width=8.5cm,angle=-90}
\caption[tbp]{\label{fig2}
A diagram which present the all possibilities for the map construction.
The wider line indicates the chosen combination.
}
\end{figure}

Up to the derivation of Eqs. (\ref{e36})-(\ref{e38}) we have passed three
stages, (a) the velocities were calculated using an exact solution and the
position $x$ can be calculated by numerical integration (Eqs. (\ref{e24})), (b)
the velocities were approximated by Eqs. (\ref{e27}) and the position $x$ by
the elementary Fresnel function, and (c) the velocities were approximated by
Eqs. (\ref{e36})-(\ref{e37}) and the position $x$ (Eq. (\ref{e38})) was
derived using elementary integration. Among all the combinations of the
solutions
for the velocities and the position $x$ we choose the combination of the
exact solution of velocities (case (a), Eqs. (\ref{e24})) and the exact
solution for the position $x$ (stage (c), Eq. (\ref{e38})). The value of $x$
was
selected from the third stage of our derivation since it enters the equation
just in the kicking term, and there it affects only slightly the phase. Thus,
one does not expect that this fact can change the typical behavior of the
particle. The above analysis is summarized in Fig. \ref{fig2}.

The radiative map is then
\begin{eqnarray}
\left({{dx} \over {dt}}\right)_{n+1} &=& 
{{\left({{dx}\over{dt}}\right)_+\cos\alpha+
\left({{dy}\over{dt}}\right)_+\sin\alpha} \over {\cosh{t\over{\tau_0}}+
\sqrt{1-\beta_+^2}\sinh{t\over{\tau_0}}}} \label{e39} \\
\left({{dy} \over {dt}}\right)_{n+1} &=& 
{{-\left({{dx}\over{dt}}\right)_+\sin\alpha+
\left({{dy}\over{dt}}\right)_+\cos\alpha} \over {\cosh{t\over{\tau_0}}+
\sqrt{1-\beta_+^2}\sinh{t\over{\tau_0}}}} \label{e40} \\
x_{n+1} &=& -{{\gamma_+} \over \Omega}\left[\left({{dy}\over{dt}}\right)_
{n+1}-\left({{dy}\over{dt}}\right)_+\right] \nonumber \\
&&-{{\gamma_+} \over {\Omega^2\tau_0}}\left[\left({{dx}\over{dt}}\right)_{n+1}
-\left({{dx}\over{dt}}\right)_+\right]+x_n \label{e41},
\end{eqnarray}
where the values immediately after the kick are as in Eqs. (\ref{e35}) and
$\alpha$ is defined in Eq. (\ref{e25}) ($\alpha \to -\alpha$ since $e=-e_0$) .
Note that for our limit ($T/\tau_0\ll 1$) there is not a great difference
between
the map given by Eqs. (\ref{e36})-(\ref{e38}) and Eqs. (\ref{e39})-(\ref{e41}).

\section{Results - weak magnetic field}

In order to obtain a web structure, there are two conditions that have to be
fulfilled. In the nonrelativistic case, the first condition is that the ratio
between the gyration frequency, $\Omega$, and the kicking time, $T$, is
a rational number, a condition which can be expressed as follows,
\begin{equation}
\label{e42}
\Omega T = 2\pi {p\over q}
\end{equation}
where $p$, and $q$, are integer numbers. Secondly, one must start from the
neighborhood of the unstable fixed point (otherwise, the particle will not
diffuse, and will not create a web structure),
\begin{eqnarray}
\left({{dx}\over{dt}}\right)_0 &=& 0 \nonumber \\
\left({{dy}\over{dt}}\right)_0 &\cong& (2n+1){{\pi\Omega}\over k}, \label{e43}
\end{eqnarray}
where $n$ is an integer number. The symmetry of the web is determined by
$p$ and $q$. If, for example $p=1$ and $q=4$, the particle is kicked four times
during one cycle, and thus, the symmetry of the web will be a four symmetry
\cite{Zaslavskii86}.

In the relativistic case \cite{Longcope87}, the above conditions are slightly
different, because of the additional factor $\sqrt{1-\beta_0^2}$ which
multiplies the $\Omega T$ term. However, if the initial velocities are small,
i.e. $v_0 \ll c$, the additional factor is close to 1, and thus the
structure of the web should not change. In the radiative and relativistic case,
as well as the non-radiative relativistic case, conditions (\ref{e42}) and
(\ref{e43}) become,
\begin{eqnarray}
\left({{dx}\over{dt}}\right)_0 &=& 0 \nonumber \\
\left({{dy}\over{dt}}\right)_0 &\cong&
(2n+1)\sqrt{1-\beta_0^2}{{\pi\Omega}\over k}. \label{e44}
\end{eqnarray}
For sufficiently large initial velocities the above conditions do not hold
anymore, and the web structure is not observed.

In this section we will compare (qualitatively; a quantitative treatment for
the diffusion rate will be studied elsewhere) the diffusion
of the non-radiative particle and the radiative particle. Intuitively, one
would expect that a radiative particle will diffuse more slowly than a
non-radiative one, since the radiation effects act like friction, and are
thus expected
to ``stop'' the particle. However, this naive expectation is not true, as we
will demonstrate below.

\begin{figure}
\psfig{figure=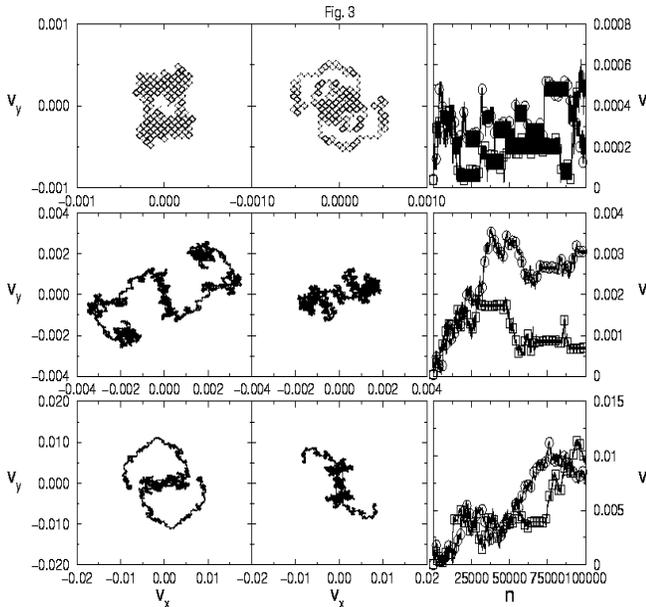,height=8cm,width=8.5cm,angle=-90}
\caption[]{\label{fig3}
Different types of diffusion behavior as described in the context.
}
\end{figure}

In order to investigate the above assumption, we have chosen a four symmetry
structure ($q=4$). We used the same initial conditions for all cases; just
the kicking strength $K$ has been changed. The parameters values and the 
initial conditions were
$B=10$, $v_{x,0}=0$, $v_{y,0}=4\times10^{-5}$, $k=10$, and the number of
iterations was $N=10^6$. In Fig. \ref{fig3}, we present the results as
follows: the
first column is the non-radiative case, the second column is the radiative
case, and the third column is the total velocity versus the iteration number
(the circles indicates the non-radiative case and the squares the radiative
one).
In the first row $K=1\times 10^{-5}$, in the second $K=2\times 10^{-5}$, and
in the third $K=4\times 10^{-5}$.

As can be seen clearly from Fig. \ref{fig3}, the web structure is valid for
the radiative case (top panel, second column). The web structure is also exists
in the other cases; the web cells are very small and they are difficult to see.
Although the radiative particle and the non-radiative particle start from the
same initial conditions they behave differently.

In some cases the diffusion rate of the radiative case is larger than the
diffusion rate of the non-radiative case, for example, in the top and the
bottom
panels after $9\times 10^5$ iterations the radiative particle reaches a higher
velocity than the non-radiative one. On the other hand, in the middle panel
the radiative particle is slower then non-radiative one. Obviously, it is
impossible to draw any general conclusion for the question of whether
a non-radiative particle is faster then the radiative one. For this a more
systematic approach should be carried out. This would be beyond the scope
of the present paper and will be considered elsewhere.

\begin{figure}
\psfig{figure=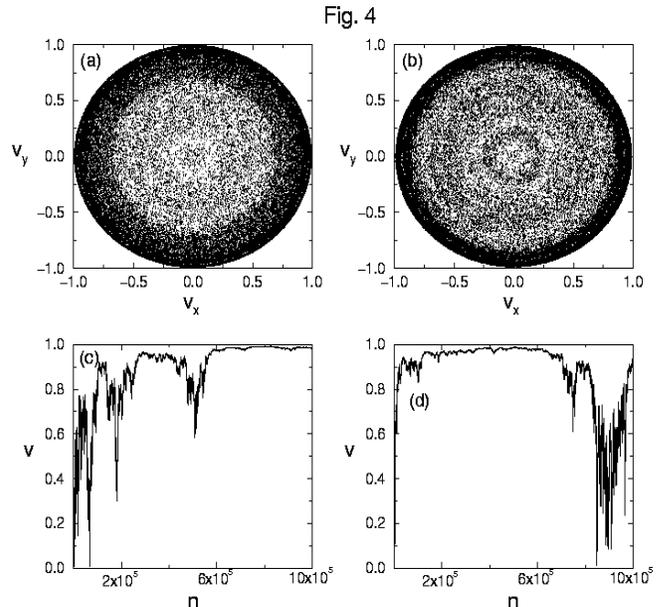,height=8cm,width=8.5cm,angle=-90}
\caption[]{\label{fig4}
The behavior of a particle with high velocity. (a) the non-radiative map,
(b) the radiative map, (c) the total velocity of the non-radiative case,
and (d) the total velocity of the radiative case.
}
\end{figure}

When the velocity of the particle reaches close to the velocity of light it
actually stops growing
(according to evolution in the time $t$) although it increases its energy. In
Fig. \ref{fig4} we show an example for that behavior. The parameters values
are : $B=10$, $v_{x,0}=0$, $v_{y,0}=1\times10^{-5}$, $k=10$, $K=0.01$, and the
number
of iterations was $N=1\times 10^6$ (every $10^{\rm th}$ iteration
was plotted). Fig. \ref{fig4}a and Fig. \ref{fig4}c are the non-radiative
cases, while Fig. \ref{fig4}b and Fig. \ref{fig4}d are the radiative cases;
in Fig. \ref{fig4}a and Fig. \ref{fig4}b the map of
$v_x$ versus $v_y$ is presented, while in Fig. \ref{fig4}c and Fig.
\ref{fig4}d the total
velocity $v$ versus the iteration number is plotted (every $10^{\it th}$
iteration was plotted). As seen from Fig. \ref{fig4}, the particle
accelerates quickly to high velocity,
and most of the time stays with this high velocity, approaching light
velocity asymptotically, while increasing its energy to infinity.

\section{Derivation of the map - strong magnetic field}

\subsection{Introductory remarks}

In the preceding section we discussed the realistic case of weak magnetic
field. In this section we will consider the other limit, i.e., the strong
magnetic field limit. Despite of the difficulties in producing a strong,
uniform, and stable magnetic field, one can assume that such fields do exist
near astronomical objects, such as, neutron stars.

When increasing the magnetic field strength the radiation effects become more
dominant; for strong enough magnetic field the particle loses most of its
energy before the next kicking. Then, the kick of the electric field ``pump''
energy into the particle, which again radiate most of this energy before the
next kick. Obviously, if the kicking strength, $K$, is small, the particle
will lose its energy in an exponential way, and will spiral in to zero
velocity (in the velocity phase space). On the other hand, if $K$ is strong
enough, the energy gain due to the kicking is larger than energy loss due the
radiation.  In this case, the particle will increase its energy (not
necessarily due to Arnol'd diffusion), and the velocity phase space will be
limited.

The interesting phenomena occur when the energy loss due to the radiation
is approximately equal to the pumping energy due to the kicking of the electric
field. In that case one can expect that : a) the velocity of the particle
will decrease to zero, b) the particle behavior (in the velocities phase space)
is a chaotic behavior, c) the particle will diffusion (while increasing
its velocity) through the web like structure, or d) the particle will increase
its velocity in a stochastic way, not through a web like structure.

\subsection{The approximated solution for the strong magnetic field case}

For strong magnetic field, we assume $T\gg\tau_0$. In this case, it is more 
convenient to write Eq. (\ref{e23}) as
\begin{equation}
\label{e45}
\beta_0 \cosh({t \over {\tau_0}}-C_1) = {1 \over 2} e^{{t \over {\tau_0}}}
\bigl(1+\sqrt{1-\beta_0^2}\bigr)\bigl(1+\xi^2\bigr),
\end{equation}
where
\begin{equation}
\label{e46}
\xi = e^{-{t \over {\tau_0}}}{\beta_0 \over {1+\sqrt{1-\beta_0^2}}}.
\end{equation}
Eq. (\ref{e25}) can be written as,
\begin{equation}
\label{e47}
\alpha = \Omega t +\Omega\tau_0\ln\biggl({{1+\sqrt{1-\beta_0^2}} \over 2}
\biggr)+\Omega\tau_0\ln(1+\xi^2).
\end{equation}
When $T\gg\tau_0$, $\xi\ll 1$ and in the first approximation, $\xi$ is
negligible. Under those circumstances, Eqs. (\ref{e24}) become,
\begin{eqnarray}
{{dx} \over {dt}}=&&{2 \over {1+\sqrt{1-\beta_0^2}}}e^{-{t \over {\tau_0}}}
\times \nonumber \\
&& \biggl[\biggl({{dx} \over {dt}}\biggr)_0\cos(\Omega t+\phi_0)-
\biggl({{dy} \over {dt}}\biggr)_0\sin(\Omega t+\phi_0)\biggr] \label{e48} \\
{{dy} \over {dt}}=&&{2 \over {1+\sqrt{1-\beta_0^2}}}e^{-{t \over {\tau_0}}}
\times \nonumber \\
&& \biggl[\biggl({{dx} \over {dt}}\biggr)_0\sin(\Omega t+\phi_0)+
\biggl({{dy} \over {dt}}\biggr)_0\cos(\Omega t+\phi_0) \biggr] \label{e49}
\end{eqnarray}
where
\begin{equation}
\label{e50}
\phi_0 = \Omega \tau_0 \ln \biggl({{1+\sqrt{1-\beta_0^2}} \over 2} \biggr).
\end{equation}
In the present case ($T\gg\tau_0$) $\phi_0\ll 0$, since $\Omega T \sim 1$
(see Eq. (\ref{e42})).

Integration of Eq. (\ref{e48}) yields,
\begin{eqnarray} 
\label{e51}
x(t) = {{\tau_0}\over{1+\Omega^2\tau_0^2}}\biggl[&&-\biggl({{dx} \over {dt}}
\biggr)(t) +\biggl({{dx} \over {dt}}\biggr)_0 + \nonumber \\
&&\Omega\tau_0\biggl(\biggl({{dy} \over
{dt}}\biggr)(t)-\biggl({{dy} \over {dt}}\biggr)_0\biggr)\biggr]+x_0.
\end{eqnarray}
We now  return to the initial charge $e=-e_0$ and thus change
$\Omega \to -\Omega$ and $\alpha \to -\alpha$. The discussions from Sec.
\ref{s4c} are valid also here; Eqs. (\ref{e39})-(\ref{e40}) remain the same and
Eq. (\ref{e41}) should be replaced by
\begin{eqnarray}
\label{e52}
x_{n+1} = -{{\tau_0}\over{1+\Omega^2\tau_0^2}}&&\biggl[\biggl({{dx} \over {dt}}
\biggr)_{n+1}
-\biggl({{dx} \over {dt}}\biggr)_++ \nonumber \\
&& \Omega\tau_0\biggl(\biggl({{dy} \over
{dt}}\biggr)_{n+1}-\biggl({{dy} \over {dt}}\biggr)_+\biggr)\biggr]+x_n.
\end{eqnarray}
Note that since the velocities in Eq. (\ref{e52}) are limited by one, and since
$\Omega\tau_0\ll 1$ and $\tau_0\ll 1$ the difference $x_{n+1}-x_n$ will be a
very small number\footnote{In order to get some impression about the size of
the magnetic field, consider the following values, $\tau_0\approx
5\times 10^8/ B^2$, $\Omega\tau_0\approx 9\times 10^{15}/B$, $T/\tau_0\approx
7B\times 10^{-16}$. Since the term $T/\tau_0$ enters in exponential
functions,  $T/\tau_0 \ge 3{1\over 2}$ is sufficient in our approximation.
Thus the magnetic field should be $B \ge 5\times 10^{11}{\rm T}= 5\times
10^{15}{\rm\, gauss}$.}.

In order to observe some interesting behavior the effective kicking should be
strong enough. The increments in the phase are very small number since
 $x_{n+1}-x_n=\order(\tau_0 v_0)$ when $B\gg 1$. Thus, it is necessary
to choose sufficiently large $k$ in the kicking
function, $f(x_n)=K\sin{kx_n}$ (Eqs. (\ref{e34})-(\ref{e35})),
\begin{equation}
\label{e53}
k=\order\bigg({1 \over {\tau_0 v_0}} \bigg).
\end{equation}
In our calculations, we used
\begin{equation}
\label{e54}
k={\pi \over {\tau_0 v_0}}.
\end{equation}
Obviously, the behavior also depends on the kicking strength, $K$. For small
$kx$ the kicking approximated by $f(x)=K\sin\,kx \approx Kkx$. Thus, in some
cases, large $K$ is equivalent to large $k$.

\section{Results - strong magnetic field}

In order to analyze the parameter space, one should first determine the most
important parameters in the system. In our case the parameters are the kicking
strength, $K$, and the magnetic field strength, $B$. The initial conditions,
$v_{x,0}$, $v_{y,0}$, $x_0$, were chosen to be much smaller than the light
velocity $c$, since in this regime the relativistic effects are
negligible, and
the radiation effects are isolated from the relativistic effects. However,
in spite of those small initial conditions, one can easily determine what would
happen if those initial conditions were in the order of the light velocity;
for large enough kicking strength, $K$, the particle will accelerate to
that regime.

Generally, a parameter space mapping requires a calculation of Lyapunov
exponents \cite{{Barreto97},{Banerjee98}}. However, in our case, because of
the decaying terms in Eqs.
(\ref{e39})-(\ref{e40}), we are interesting in the decaying regions. For this,
the standard deviation,
\begin{equation}
\label{e55}
\Delta v = \sqrt{<v^2>-<v>^2},
\end{equation}
is considered (after a sufficiently large number of iterations when the motion
converges to its
stable behavior). In the case of decaying velocities, $v \to 0$, and thus,
$\Delta v \to 0$. In the most cases of periodic motion $v={\rm const}$ and
thus, $\Delta v =0$. As will be shown in the following, there exists a
bifurcation
behavior in the system, and since the gap between bifurcation points
decreases exponentially, the period one region is much larger than the other
periods (the exponential decrease start from the point between period one and
period two, and do not include the period one itself). Moreover, since it is
very difficult to balance the kicking strength, $K$, with the magnetic field
strength, $B$, in such a way that the pumping energy due to the kick will be
radiated before the next kick, just a small part of the parameter space will 
lead to a quasiperiodic or to a
periodic motion. The dominant part of the parameter phase
space should be either chaotic or stochastic, and thus should be indicated by
a larger value of $\Delta v$. A high value of $\Delta v$ simply means that the
basin of attraction (if it exists) is larger.

\begin{figure}
\psfig{figure=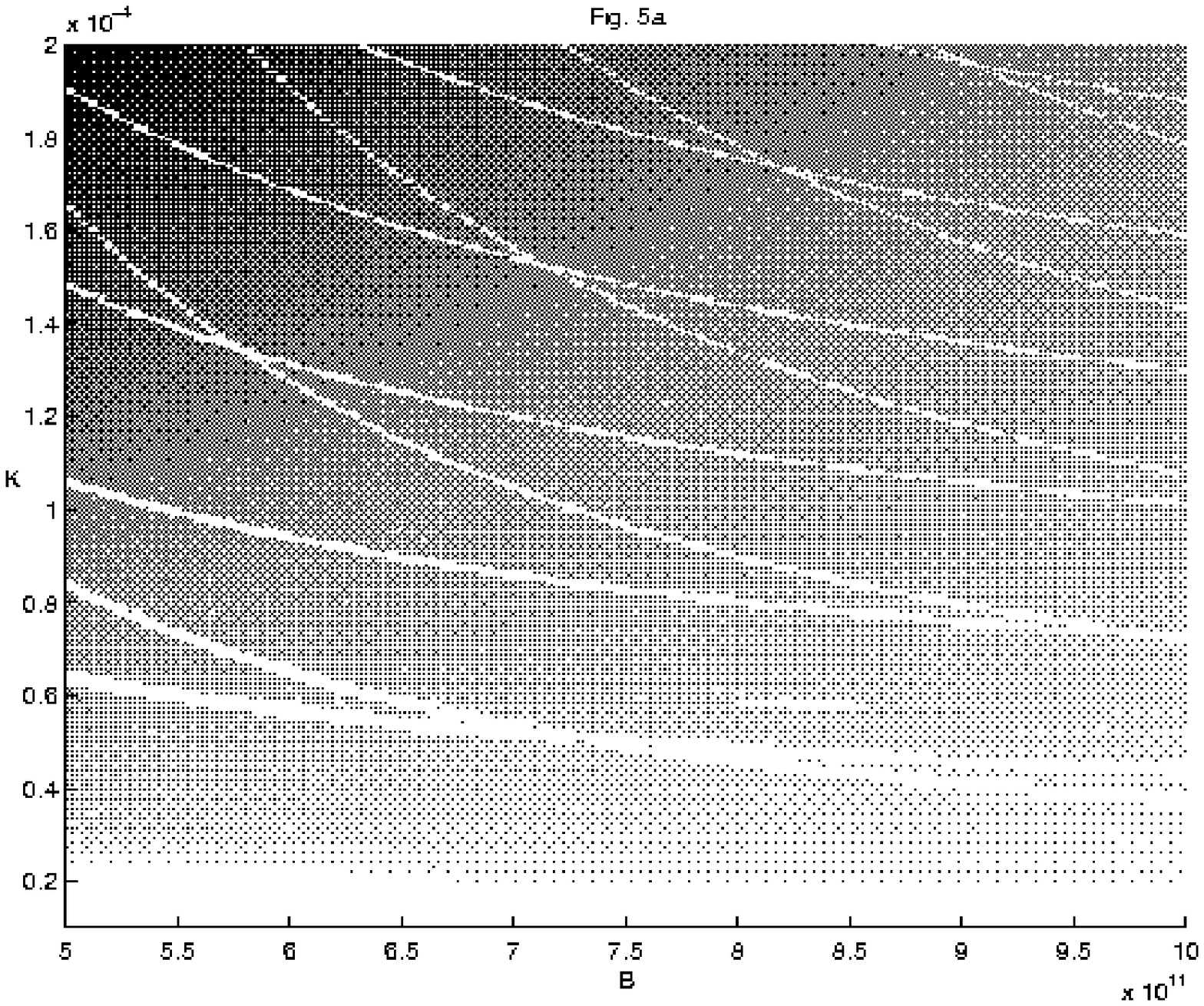,height=8cm,width=8.5cm}
\psfig{figure=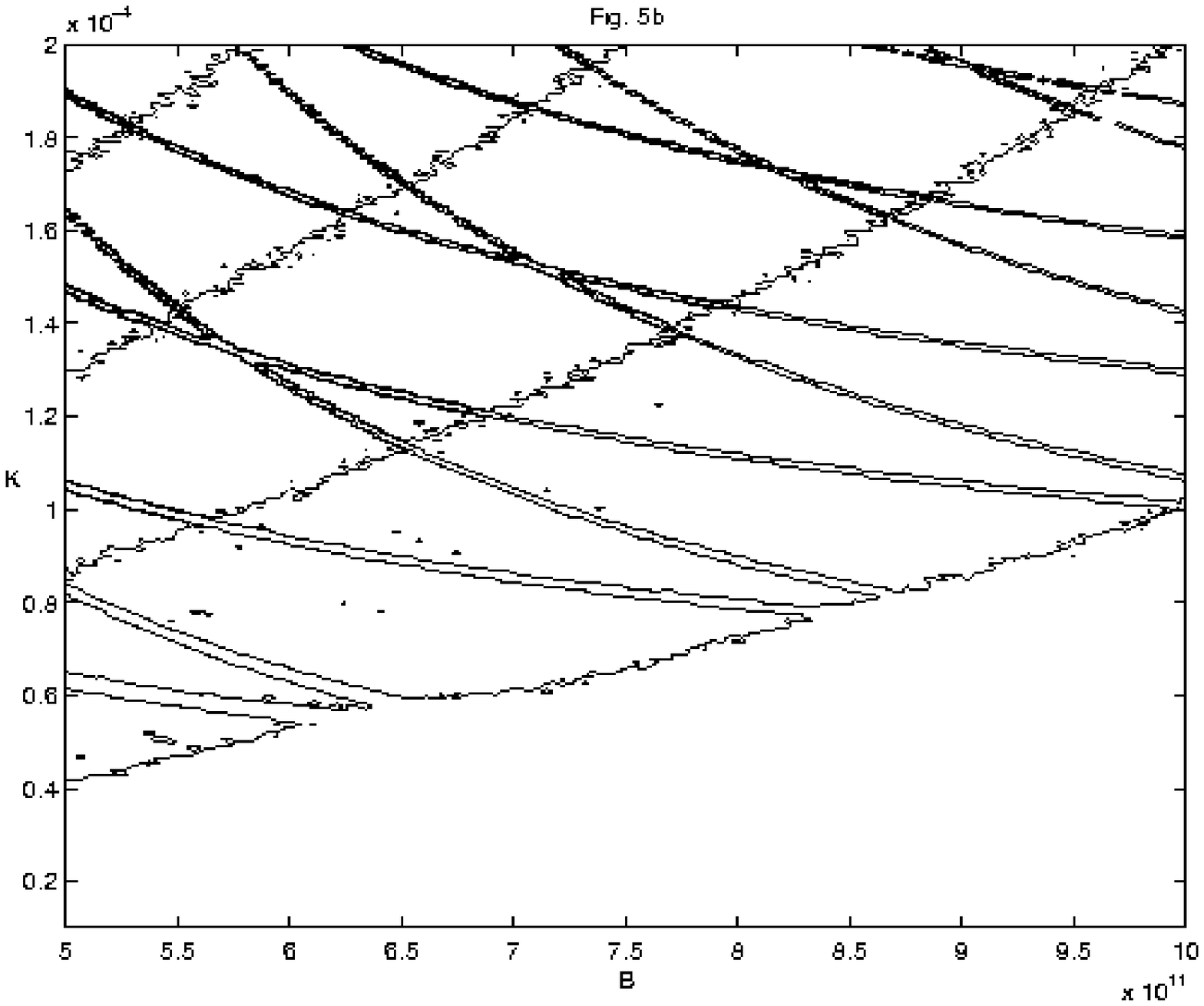,height=8cm,width=8.5cm}
\caption[]{\label{fig5}
The parameter space, $B:K$, mapping. (a) A density plot of the parameter space.
The darker regions indicate high velocity standard deviation. (b) Contour plot
of (a).}
\end{figure}

In Fig. \ref{fig5}a we present a density plot of the parameter phase space (the
height is $\Delta v$). The darker regions have higher $\Delta v$. For
small $K$ the velocity decays to zero. For larger $K$ the basin of attraction
is larger, and one expect to find a more complex behavior. However, there
exist {\it stability lines} which reflect that the particle either loses its
energy or moves in a periodic motion. The stability lines seems to be
continuous lines with well defined shape (the small discontinuities can be
related to a high period level of the orbit; in this case $\Delta v \not \to
0$ but $\Delta v$ will be relatively small). Even for large $K$ values, in
which the particle
is kicked very strongly, it can move in a periodic way. However,
those lines accumulate just a minor part of the parameter phase space, and
thus one can expect to find a chaotic behavior near the neighborhood of an
arbitrary point in parameter phase space \cite{Barreto97}. Such a parameter
phase space gives an indication of a robust chaotic system \cite{Banerjee98}
with three positive
Lyapunov exponents\footnote{According to the conjecture in Ref.
\cite{Barreto97}, if the number of
positive Lyapunov exponents is larger than the number of the relevant
parameters, the chaotic behavior is ``robust''; for an arbitrary
point in the parameter space one expects to find a chaotic behavior in the
neighborhood of this point. When the number of positive Lyapunov exponents is
less or equal to the number of relevant parameters, the chaotic behavior is
less
strong; in the neighborhood of an arbitrary point in the parameter space it is
difficult to find a chaotic behavior. In our case, the map is a three 
dimensional
map, and thus the fractal dimension must be less or equal to four (one should
take into account the time, $t$; it did not enter explicitly in the map). In 
the
more realistic case the map never covers the entire phase space and thus the
fractal dimension will be less then four. Thus, the number of positive
Lyapunov exponents should be three.}. Fig. \ref{fig5}b shows a contour plot of
Fig. \ref{fig5}a.

\begin{figure}
\psfig{figure=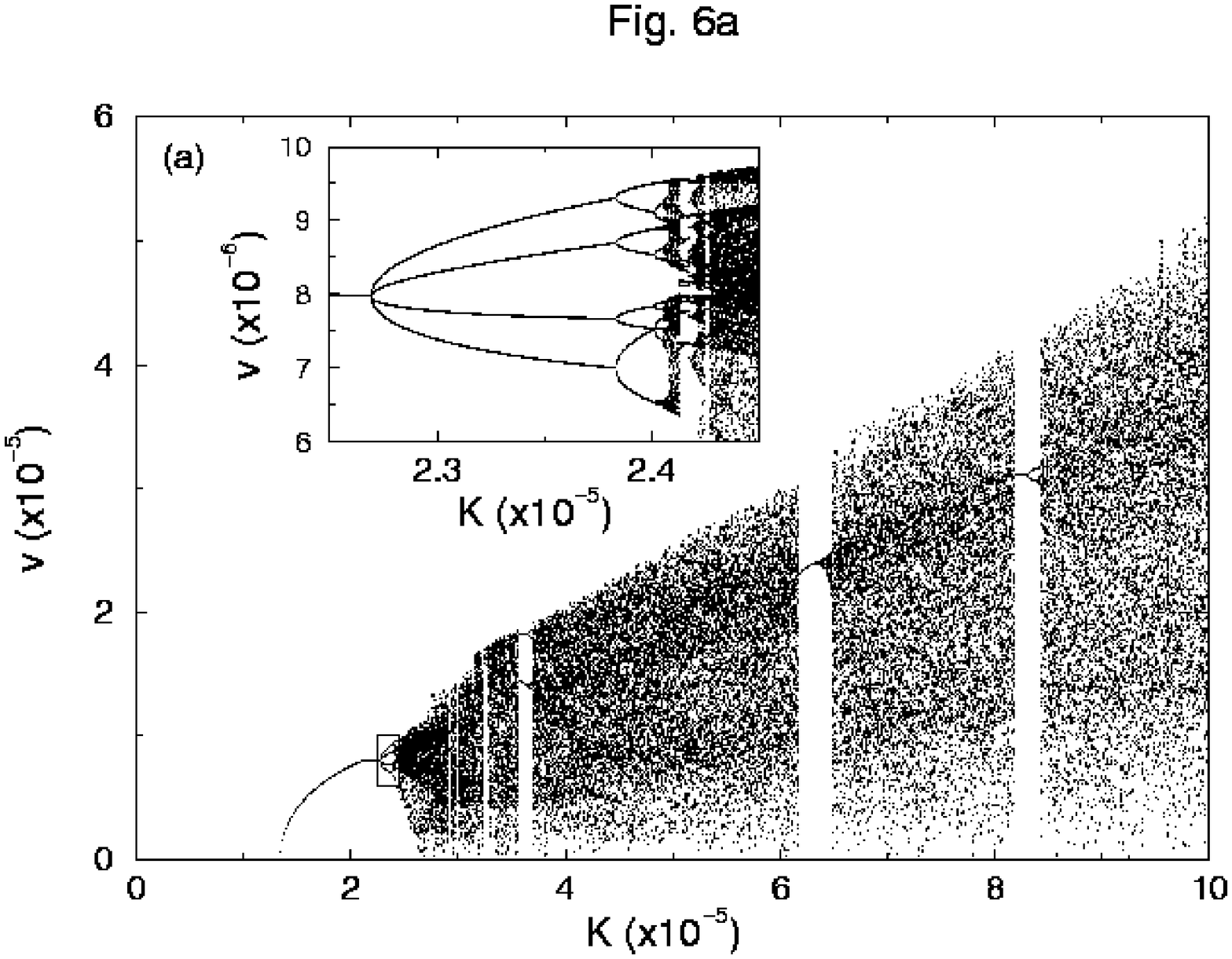,height=8cm,width=8.5cm,angle=-90}
\psfig{figure=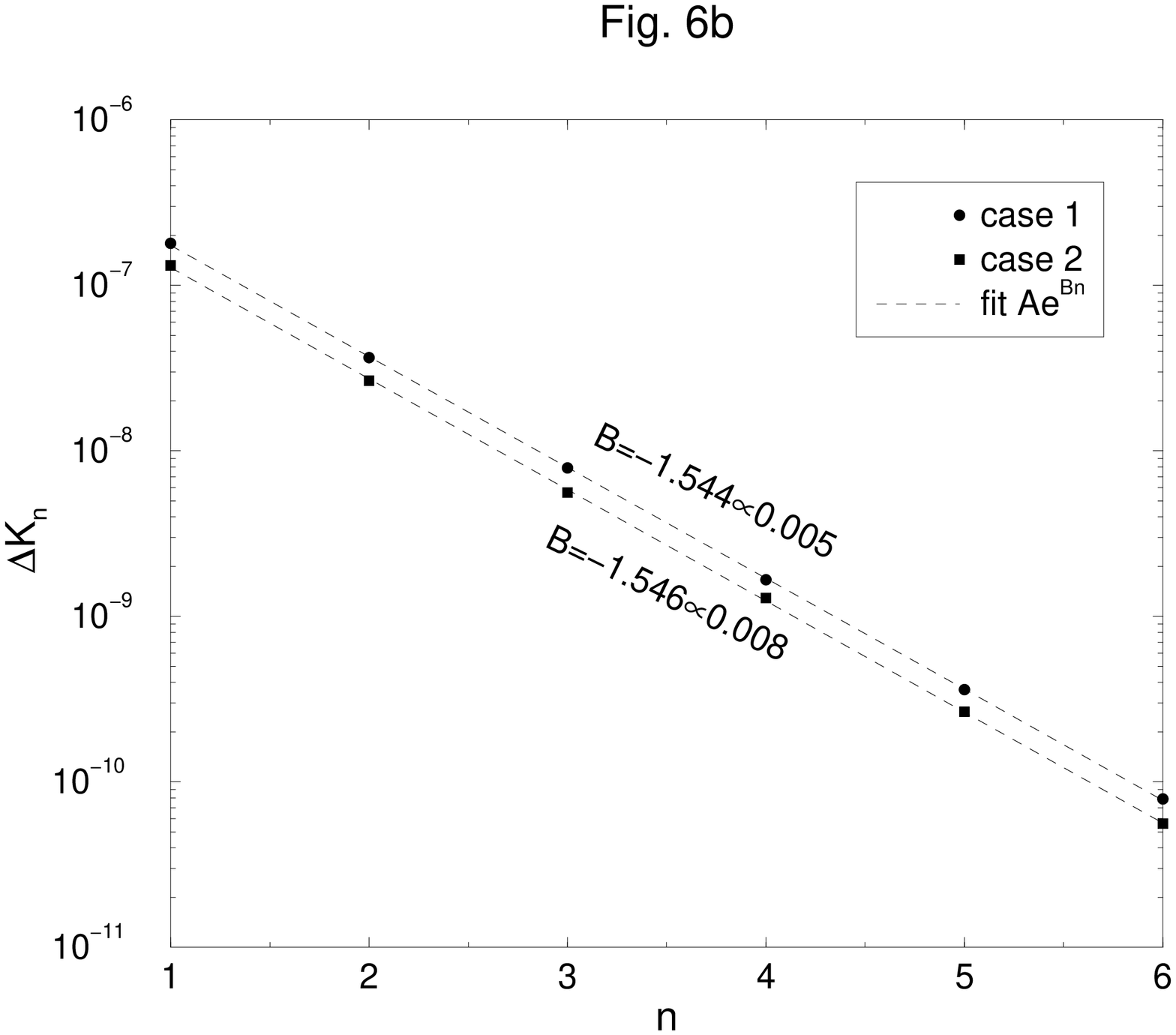,height=8cm,width=8.5cm,angle=-90}
\caption[]{\label{fig6}
The bifurcation behavior of the radiative map (in the strong magnetic field
limit). (a) The velocity $v$ versus the kicking strength $K$. The inset shows
an enlargement of the indicated rectangle. (b) The spacing between the
consecutive bifurcation points, $\Delta K_n$, versus the bifurcation number,
of two different cases (as mentioned in the text). The dashed line is the
exponential fit of the two cases.}
\end{figure}

Taking the a fixed magnetic field value, B, and changing the kicking strength,
$K$, will give more inset about the particle behavior. In Fig. \ref{fig6}a
we have plotted a bifurcation map using $K$ as the relevant parameter. For each
$K$ value we have iterate the map for several thousand of iterations and
plotted
just the last few hundred of points (this have done to make sure that the
particle was convergent to its basin of attraction). For small $K$ values, the
particle do not kicked strong enough, and it loses its energy exponentially to
zero. For higher $K$ values ($K=1.35\times 10^{-5}-2.25\times 10^{-5}$) the
particle has a constant velocity. Then, a series of bifurcation points appears,
leading to the chaotic region (the enlargement of the solid rectangle is shown
in the inset, where the bifurcation points can be clearly seen). The chaotic
region which follows afterwards ends with other period one points,
followed by another set of bifurcation points, and so on. The regular regions
of Fig. \ref{fig6}a are parts of the stability line of Fig. \ref{fig5}. The
maximum velocity that the particle reaches seems to grow linearly.

The series of bifurcation points is connected to the universal constant
$\delta\approx 4.6692$, which was discovered by Feigenbaum \cite{Feigenbaum78}.
The bifurcation
phenomenon is one of the signs of chaotic behavior. In Fig. \ref{fig6}b we
present the exponential decaying behavior of the displacement between
bifurcation points, $\Delta K_n$, in the inset of Fig. \ref{fig6}a (case 1)
as well as, for another bifurcation region ($K=8.25\times 10^{-5}-8.45\times
10^{-5}$ denoted as case 2). As can be seen, in the two cases the exponent is
almost the same, $B\approx -1.545$. The connection between the exponent $B$ and
universal constant $\delta$ is given by,
\begin{equation}
\label{e56}
{{\Delta K_n} \over {\Delta K_{n+1}}}=e^{-B}=e^{1.545}\approx 4.688 \approx
\delta= 4.6692
\end{equation}
According to Feigenbaum \cite{Feigenbaum78} the convergence to the universal
constant $\delta$ should be when $n \to \infty$ (in $\Delta K_n$). However, in
the present case the convergence is rapid; the fitting of the
exponential function is from the second bifurcation point.

\begin{figure}
\psfig{figure=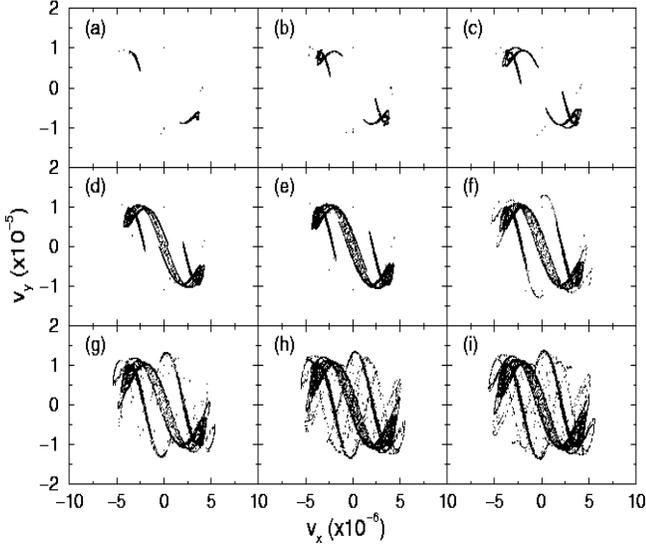,height=8cm,width=8.5cm,angle=-90}
\caption[]{\label{fig7}
The chaotic attractors of increasing $K$ values: (a) $K=2.45\times 10^{-5}$, 
(b) $K=2.5\times 10^{-5}$, (c) $K=2.55\times 10^{-5}$, 
(d) $K=2.6\times 10^{-5}$, (e) $K=2.65\times 10^{-5}$, 
(f) $K=2.7\times 10^{-5}$, (g) $K=2.75\times 10^{-5}$,  
(h) $K=2.8\times 10^{-5}$, and (i) $K=2.85\times 10^{-5}$.
For larger $K$ values the
attractor occupied larger area of the velocity space.}
\end{figure}

The evolution of the chaotic behavior as a function of the kicking strength,
$K$, is shown in Fig. \ref{fig7} (the initial conditions are $V_{x,0}=0$,
$V_{y,0}=10^{-5}$ and the parameters values are $B=5\times 10^{11}{\rm T}$,
$q=4$; each figure contains 10000 points).
In Fig. \ref{fig7}a there are two separate
regions which become closer for increasing $K$, and finally join to one
region (Fig. \ref{fig7}d). Moreover, the attractor occupies more area of the
velocity space. The complexity of the attractor seems to be richer for larger
$K$ values. However, the fractal dimension of the attractors of Fig. \ref{fig7}
do not increase monotonically \cite{Ashkenazy}. The fractal dimension of
Fig. \ref{fig7} is
summarized in Table \ref{table1}. We have calculated the fractal dimension,
$D_2$, using two different methods, the generalized information dimension
(column 2 in Table \ref{table1}) and the generalized correlation dimension
(column 3 in Table \ref{table1}). The
fact that the value of the fractal dimension is not an integer number is
another indication for the chaotic behavior of the system. The higher the
fractal dimension is, the greater is the complex behavior. Additionally, the
non-integer fractal dimension indicates at least one positive Lyapunov 
exponent,
(using the above arguments, there are three positive Lyapunov exponents)
which is another indicator of chaotic behavior\footnote{It follows from
Kaplan-Yorke conjecture \cite{Kaplan} that if the fractal dimension is
a non-integer number there exists, at least, one positive Lyapunov exponent.}.

\begin{figure}
\psfig{figure=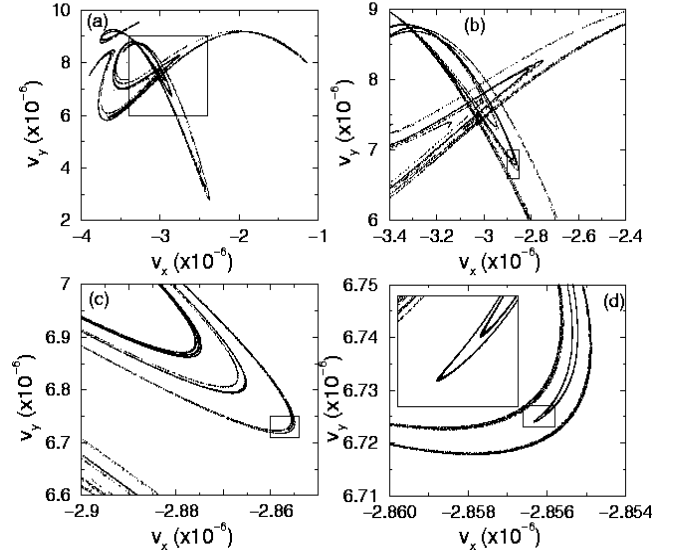,height=8cm,width=8.5cm,angle=-90}
\caption[]{\label{fig8}
The self-similarity and the folding-stretching phenomenon (starting from Fig.
\ref{fig7}b). Each figure is an enlargement of the indicated rectangle of the
previous figure.}
\end{figure}

\begin{figure}
\psfig{figure=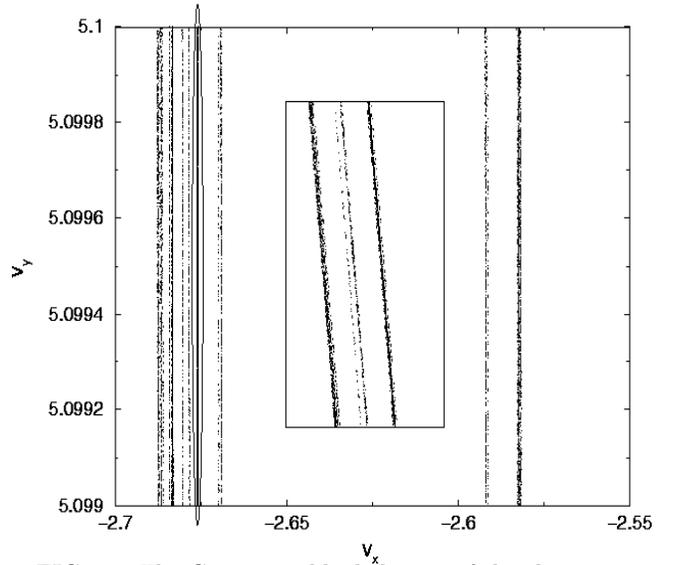,height=8cm,width=8.5cm,angle=-90}
\caption[]{\label{fig9}
The Cantor set like behavior of the chaotic map. The inset shows the
self-similarity of the attractor, keeping the Cantor set behavior of the lines.
}
\end{figure}

One of the signs of chaotic behavior is the folding-stretching phenomenon,
as well as the self-similarity of the map \cite{Guckenheimer83}.
 Fig. \ref{fig8} shows the ``inner
view'' of Fig. \ref{fig7}b. Each figure is an enlargement of the previous
indicated rectangle, and contains 10000 points. The self-similar structure is
clearly seen from Fig. \ref{fig8}c and \ref{fig8}d. The location of the
``attractor
lines'' seems to form a Cantor set which is usually observed in chaotic maps.
The folding-stretching phenomenon which is connected to the self-similar
structure can be seen, for example, in Fig. \ref{fig8}d where the solid
rectangle shows sharp turns of the maps.
The Cantor set nature of the attractor lines is demonstrated in Fig. \ref{fig9}
where we choose a horizontal cut of Fig. \ref{fig8}a. There are more dense
regions and empty regions, as expected. The inset of the figure shows that
this behavior continues at smaller scales.
\begin{table}[h]
\vbox{\offinterlineskip
\halign{\strut \vrule \hfil \quad #\quad \hfil \vrule &\vrule \hfil \quad #\quad \hfil
&\vrule \hfil \quad #\quad \hfil \vrule \cr
\noalign{\hrule}
Fig. \# & Info. D. & Corr. D. \cr
\noalign{\hrule}
\noalign{\hrule}
   a & 1.20 & 1.13\cr
\noalign{\hrule}
   b & 1.32 & 1.27\cr
\noalign{\hrule}
   c & 1.41 & 1.39\cr
\noalign{\hrule}
   d & 1.43 & 1.43\cr
\noalign{\hrule}
   e & 1.39 & 1.41\cr
\noalign{\hrule}
   f & 1.36 & 1.36\cr
\noalign{\hrule}
   g & 1.38 & 1.32\cr
\noalign{\hrule}
   h & 1.47 & 1.49\cr
\noalign{\hrule}
   i & 1.52 & 1.53\cr
\noalign{\hrule}
}}
\caption[]{\label{table1}
The fractal dimension, $D_2$, of Fig. \ref{fig7}. The first column is the
figure number, the second column is the information dimension, and the third
column is the correlation dimension.}
\end{table}

\begin{figure}
\psfig{figure=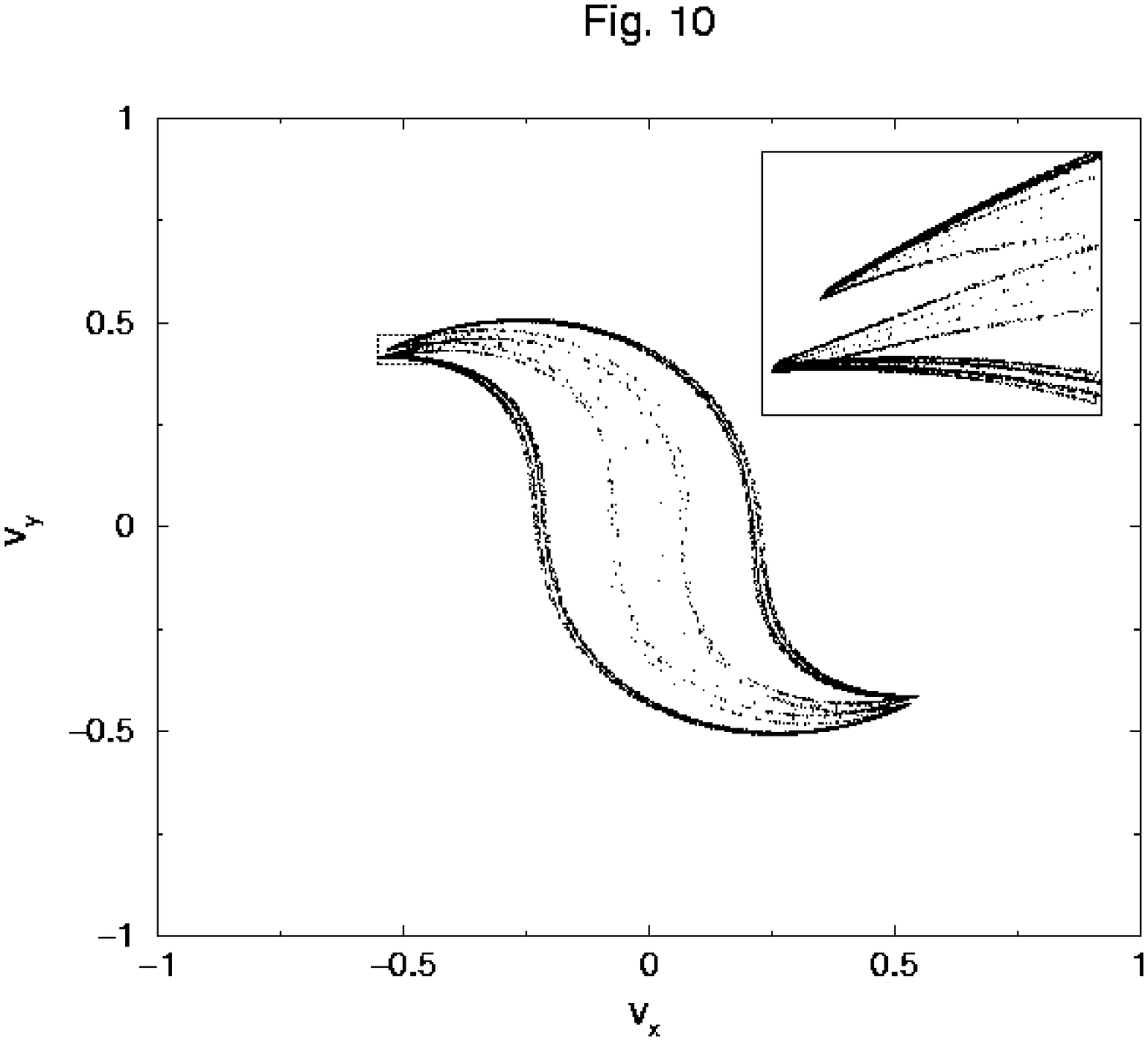,height=8cm,width=8.5cm,angle=-90}
\caption[]{\label{fig10}
The chaotic behavior of the radiative map (large $B$) for large kicking
strength.}
\end{figure}

For large kicking strength the particle increases (or decreases) its velocity
to a region which balances between the kicking and the radiation. In Fig.
\ref{fig10} an example for the limited map is shown (the parameters values are:
$B=5 \times 10^{11}$, $q=4$, $K=30$, $v_{x,0}=0$, and $v_{y,0}=0.99$).
The clear boundaries are caused by the radiation which decreases the velocity
significantly before the next kick. The kicking pushes the particle toward
 light velocity; the gap between light velocity and the attractor
boundaries is approximately the energy loss caused by the radiation. The
self-similar behavior and the strength-folding phenomenon can be seen in the
inset of Fig. \ref{fig10}.

\section{Intermediate magnetic field}

In the previous sections the analysis of the weak and the strong magnetic field
was discussed. In this section we will discuss the intermediate magnetic field
strength regime. Basically, a numerical integration of Eq. (\ref{e24}) should
be performed. However, the approximation for the weak magnetic field (Eqs.
(\ref{e39})-(\ref{e41}) is valid up to $B=10^{10}{\rm T}$ \footnote{As it
already pointed out $\tau_0\approx 5\times 10^8/ B^2$,
$\Omega\tau_0\approx 9\times 10^{15}/B$, $T/\tau_0\approx 7B\times 10^{-16}$.
Thus, the requirement $B\le 10^{14}{\rm gauss}=10^{10}{\rm T}$ will
satisfied the condition $T/\tau_0 \ll 1$.}, and thus one can avoid the
numerical integration.

\begin{figure}
\psfig{figure=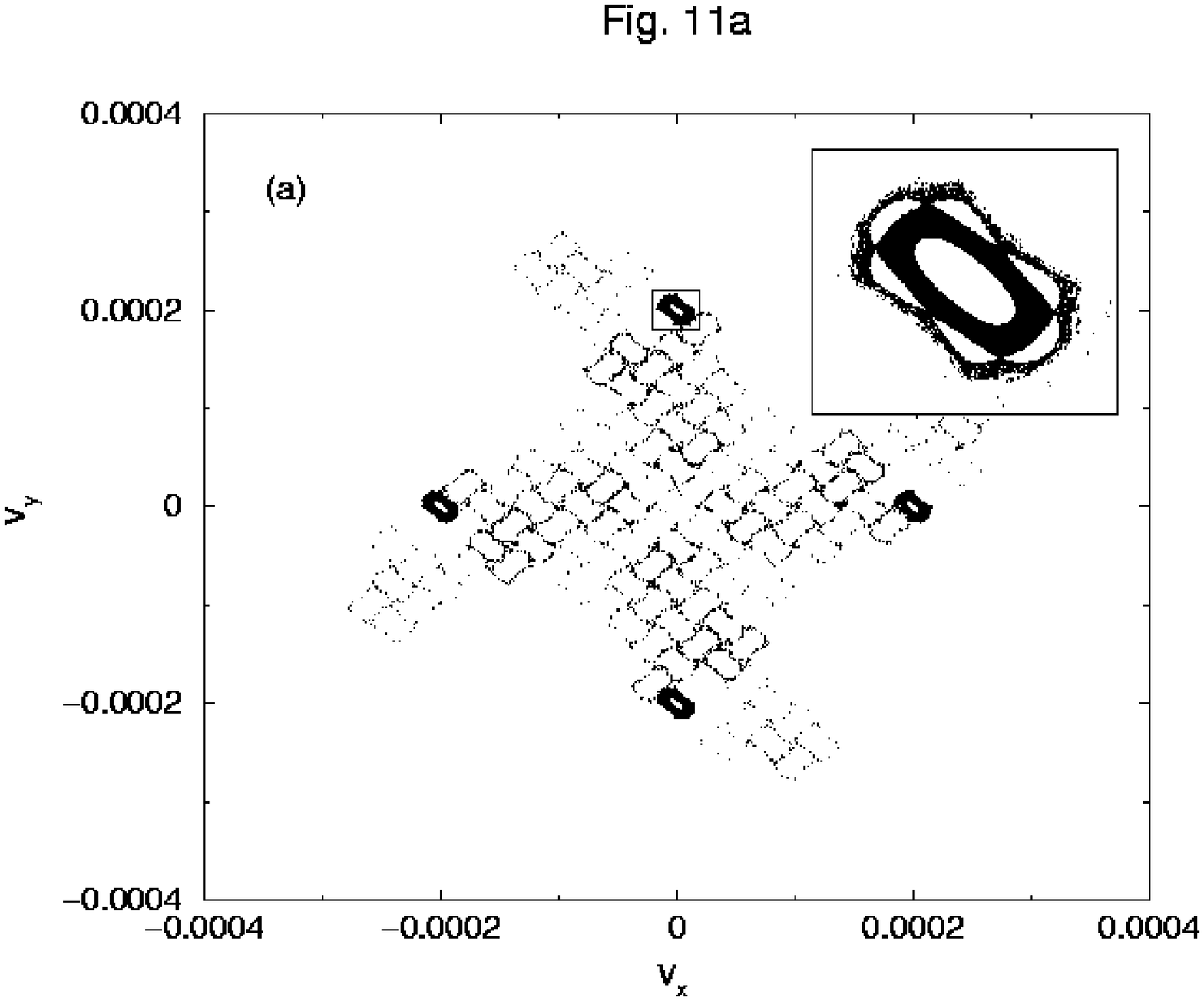,height=7.5cm,width=8.5cm,angle=-90}
\psfig{figure=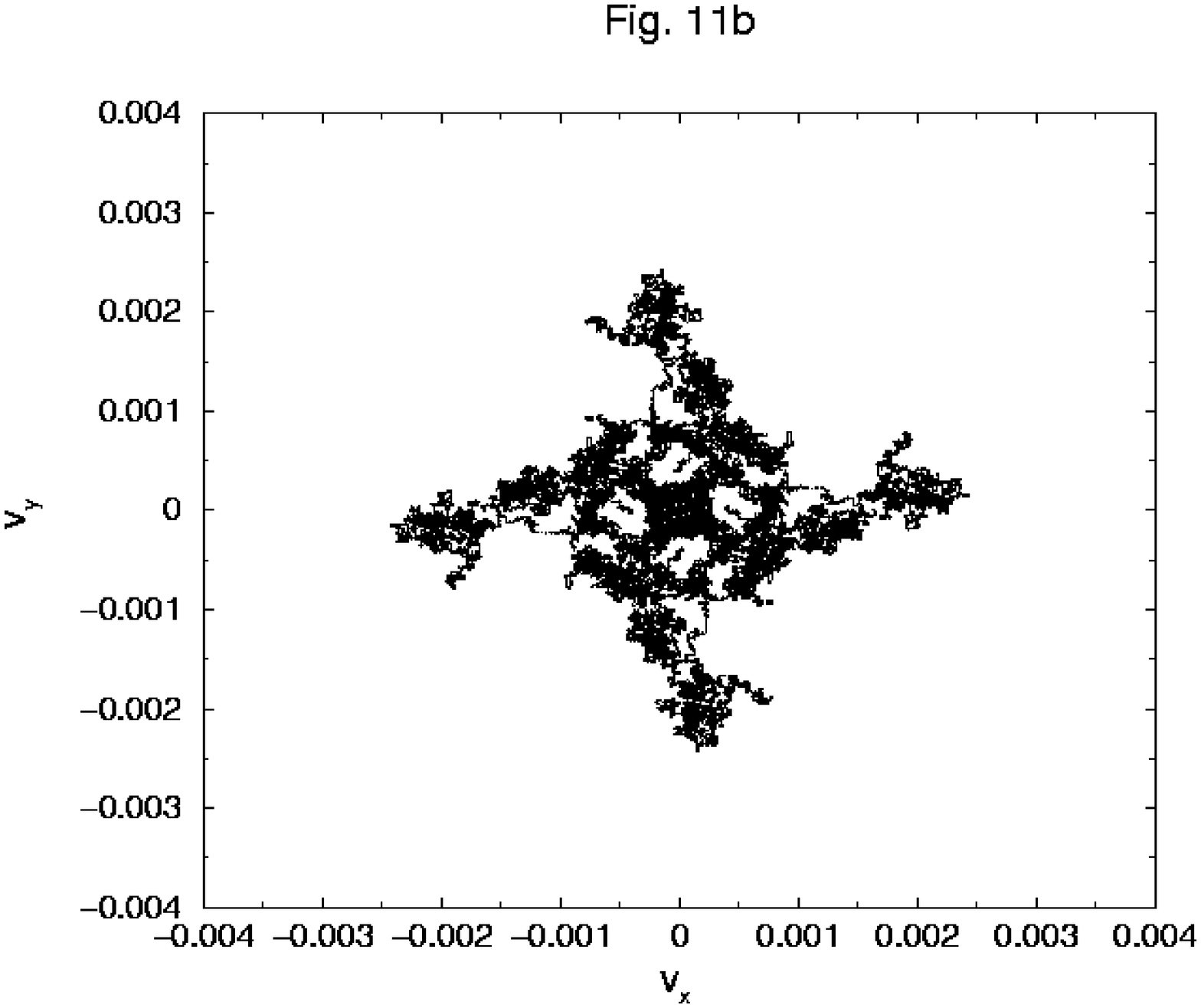,height=7.5cm,width=8.5cm,angle=-90}
\caption[]{\label{fig11}
The intermediate magnetic field regime. The parameter values are
$B=5\times 10^6{\rm T}$, $q=4$, $K=10^{-5}$, and $v_{x,0}=0$. Each figure
consist $10^5$ points. The initial velocity, $v_{y,0}$, in (a) is
$2\times 10^{-5}$ while in (b) is $1\times 10^{-5}$.}
\end{figure}

In Fig. \ref{fig11} some of the possibilities that can appear in the
intermediate magnetic field regime are presented. For
example, Fig. \ref{fig11}a shows a destroyed web like structure, i.e., the
particle started to diffuse through a web structure, but after a sufficiently
large number of iterations it either decays toward zero velocity, or stops its
acceleration and iterates in a quasiperiodic motion (see the enlargement of
the indicated rectangle in Fig. \ref{fig11}a, where the particle decays 
towards a periodic orbit). The particle
starts near an unstable fixed point (Eq. (\ref{e44})), and since the magnetic
field is relatively high its influence on the particle is larger and it
decays inside. For smaller magnetic field the
particle diffuses through the filaments of the web. In might be that, if the
initial condition were chosen closer to the unstable fixed point, the particle
would diffuse. A supported example for that claim can be seen in Fig.
\ref{fig11}b where
the particle starts from smaller velocity (half of the velocity of Fig.
\ref{fig11}a) using the same parameters values as in
Fig. \ref{fig11}a (and thus the approximation of Eq.
(\ref{e44}) is more accurate) and it diffuses to the high velocity regime
(approximately, to a ten times larger velocity).
Thus, one can safely conclude that for intermediate magnetic field the 
influence
of the unstable fixed point is somehow weaker, and it is more difficult to
observe the web structure. Or, to state it in a different way, the width
of the filament is narrower when the magnetic filed is stronger.

\section{Summary}

In the present paper we have investigated the effect of radiation on the
stochastic web. Under the restriction of small magnetic fields
($B \lesssim \order(100T)$) an iterative map was constructed. Moreover, the
effect of
radiation on the stochastic web is very small because of the small magnetic
field. Qualitatively, the non-radiative and the radiative cases have similar
web structure. Despite of the naive expectation that the diffusion rate in the
radiative case should be smaller than non-radiative case, one can find cases
in which the opposite effect is observed.

Although it seems the effect of radiation is not qualitatively
significant under
laboratory conditions, it have a strong effect in the presence of a
strong
magnetic field. Such a magnetic field could occur near or in a neuron star (or
other heavy stars) and it can cause a large radiation correction to the motion
of the particle. An iterative map was constructed for that regime also. It was
found that the web structure disappears and is replaced by rich chaotic
behavior which is demonstrated by the  non-integer fractal dimension,
bifurcation, self-similarity, and folding and stretching.
The chaotic attractor occupies a  small part of the velocity space, and
practically means that the particle can not accelerate to infinite energy, as
in the regular stochastic web.
This chaotic behavior seems to be a ``robust'' chaotic behavior (implying that
one can find chaotic behavior using slightly different parameter values).

For the intermediate magnetic field strength, the influence of the unstable
fixed point is smaller relative to a weaker magnetic field.

\section{Acknowledgements}

We would like to thank I. Dana and A. Priel at Bar Ilan University, for helpful
discussions, and S. Cohen and R. White for their interesting comments at a 
seminar
at the Princeton Plasma Physics Laboratory where some of these results
were presented.
One of us (LPH) wishes to thanks S.L. Adler for his kind hospitality
at the Institute for Advanced Study, where this work was completed.



\begin{references}
\bibitem{Ashkenazy} For details about the calculation of the generalized
information dimension and the generalized correlation dimension see
Y. Ashkenazy, ``The Use of Generalized Information Dimension in Measuring
Fractal Dimension of Time Series'', To be published in Physica A,
chao-dyn/9805001.
\end{references}
\end{document}